\documentclass[trackchanges, twocolumn]{aastex701}

\usepackage{natbib}
\usepackage{amssymb}

\submitjournal{ApJL}

\begin{document}

\title{Discovery of Optical Filaments in the North Polar Spur/eROSITA Bubble}

\author[orcid=0009-0008-7952-5702,sname='North America']{Julian Shapiro}
\affiliation{Department of Astronomy, University of California, Berkeley, CA 94720-3411, USA}

\email[show]{julianshapiro@berkeley.edu}  

\begin{abstract}

I report the discovery of optical filaments in H$\alpha$, [O\,{\sc iii}], and [S\,{\sc ii}], coincident with $144\,\mathrm{MHz}$ structure, associated with the North Polar Spur (NPS) region of the eROSITA bubbles. The optical emission is identified in the Northern Sky Narrowband Survey, and contains distinct structure and potentially elevated $\text{[S\,{\sc ii}]}/\mathrm{H}\alpha$ relative to typical high-$b$ outflows from the warm ionized medium (WIM). $\Delta$-variance and power spectrum analysis of the NPS H$\alpha$ shows a preference for filaments on the scale of lag $L\approx0.6\degr$. The optical and $144\,\mathrm{MHz}$ emission peaks in intensity in an interior region of the NPS, aligned with surrounding Galactic magnetized structures, and also shows fainter signal at the eROSITA bubble edge. The required power to produce the background-subtracted $I_{\mathrm{H}\alpha}=1.85\pm1.04\,\mathrm{R}$ emission is $P_\mathrm{req}\approx5\mathrm{-}12\times10^{41}\,\mathrm{erg\,s}^{-1}$, comfortably exceeded by AGN jet models, plausible for AGN hot accretion flow and star formation ring models, and disfavoring star formation driven winds.

\end{abstract}

\section{Introduction}

\begin{figure*}
    \centering
    \includegraphics[width=1\linewidth]{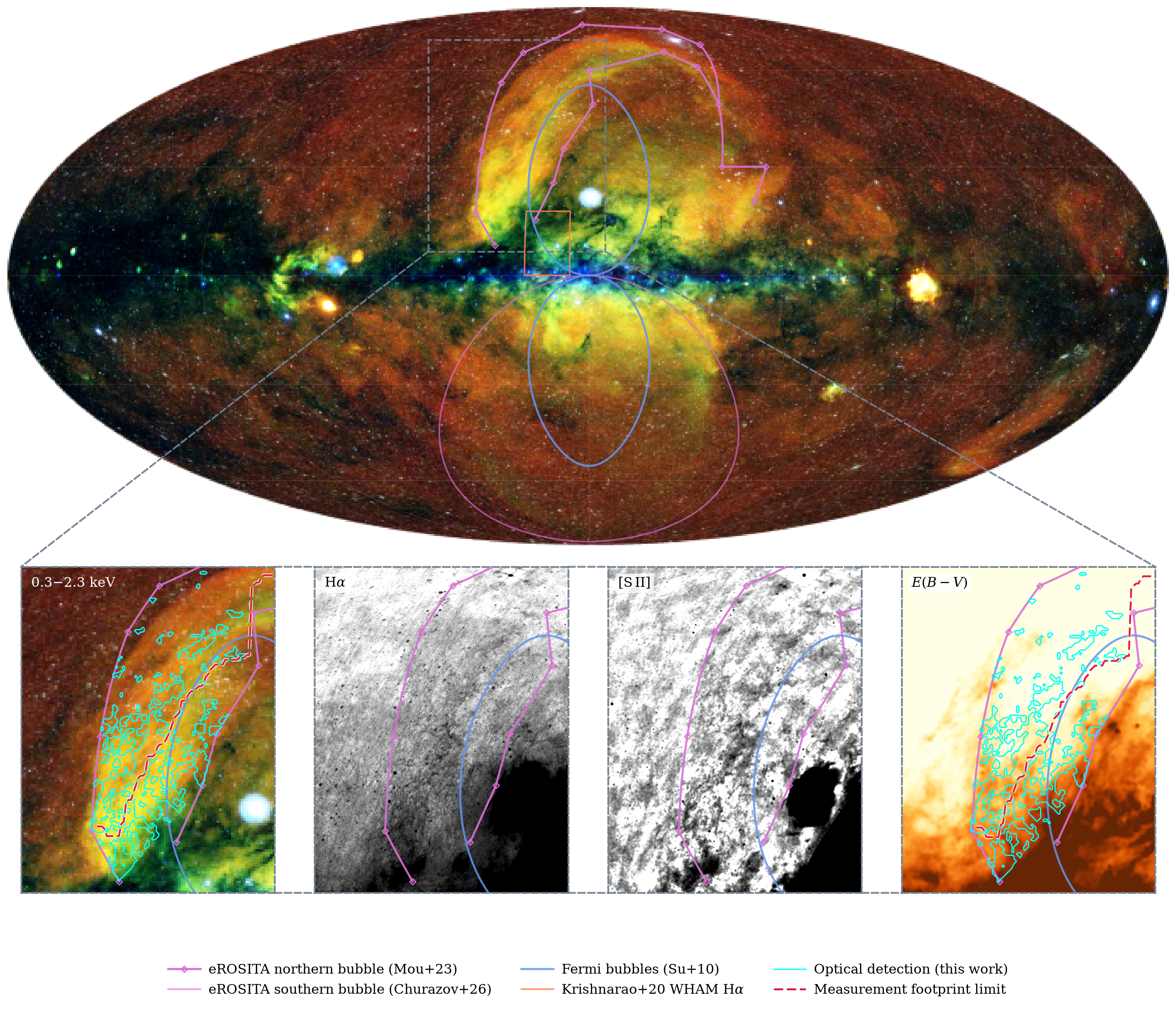}
    \caption{All-sky map in eROSITA $0.3\mathrm{-}2.3\,\mathrm{keV}$ (top; \citealt{Predehl2020}) with the geometry of the eROSITA bubbles \citep{Mou2023, churazov2026southernerositabubbleforward} and approximate extent of the Fermi bubbles \citep{Su_2010} highlighted. The orange box shows the coverage of the Fermi bubbles analyzed in WHAM H$\alpha$ by \citet{Krishnarao_2020}. 
    The bottom row displays cutouts of the North Polar Spur in eROSITA $0.3\mathrm{-}2.3\,\mathrm{keV}$, NSNS H$\alpha$, NSNS [S\,{\sc ii}], and $E(B-V)$ from \citet{Green_2019}. The cyan contours show the optical consensus mask in the eROSITA bubble, and the dashed red line marks the lower boundary of the footprint for all measurements, based on the Haslam mask overlap.}
    \label{fig:maps}
\end{figure*}

The eROSITA bubbles \citep{Predehl2020} are a pair of giant X-ray emitting shells that seem to emanate from the Galactic center, spanning over $80\degr\times80\degr$ \citep{yeung2026srgerositadiffusesoftxray}. The structure is remarkably similar to and appears to encompass the $\sim55\degr\times45\degr$ Fermi bubbles \citep{Su_2010} spanning $\sim7\,\mathrm{kpc}\times10\,\mathrm{kpc}$. These smaller structures are approximately uniform in surface brightness and demonstrate north-south symmetry, perpendicular to the Galactic plane. A coincident ``haze'' signal was identified earlier in the WMAP microwave images \citep{Finkbeiner_2004}. The hard Fermi spectrum was analyzed between $100\,\mathrm{MeV}$ and $500\,\mathrm{GeV}$ by \citet{Ackermann_2014}, finding a far lower spectral index than typical Galactic inverse-Compton emission or pion-decay emission.

There is compelling evidence for the Fermi bubbles originating from the Galactic center (GC) \citep{Su_2010}. This is often attributed to AGN activity from Sgr A*, though models are debated. Hydrodynamical simulations show consistency with a jet-driven bubble model, occurring $1\mathrm{-}6\,\mathrm{Myr}$ ago for $\sim0.1\mathrm{-}0.5\,\mathrm{Myr}$ \citep{Guo_2012, Yang_2012, 10.1093/mnras/stt1772, Yang_2017}. The diffusion of cosmic rays along the magnetic field lines surrounding the bubble may produce the observed sharp edges in $\gamma$-ray observations \citep{Yang_2012}. \citet{10.1093/mnras/sty1533} analyze the ROSAT data at differing distances from the edges, finding that the X-ray brightness declines significantly as it is crossed, supporting a $\mathcal{M}\sim4$ forward shock. Further models indicate that the jet-powered forward shocks imply collimated injection nearly perpendicular to the Galactic plane (e.g. \citealt{Zhang_2020, 10.1093/mnras/stac1084}).

Alternatively, the Fermi bubbles are proposed to be a result of a lower luminosity hot accretion flow over $\sim10\,\mathrm{Myr}$ \citep{10.1111/j.1365-2966.2012.21250.x, Mou_2014, Mou_2015, 10.1093/mnras/stx314, 10.1093/mnras/stac3312}. Under this interpretation, the wide-angle outflows from the AGN interact with the ISM and the Central Molecular Zone to form the teardrop shapes of the bubbles (e.g. \citealt{10.1111/j.1365-2966.2012.21250.x}).

Observations of [C\,{\sc iv}]/[C\,{\sc ii}] and [Si\,{\sc iv}]/[Si\,{\sc ii}] UV absorption line ratios, as well as H$\alpha$ emission of the Magellanic Stream, have been used to propose Seyfert flare activity close to the Eddington limit \citep{Bland-Hawthorn_2013, Bland-Hawthorn_2019}. However, the flare interpretation overpredicts the observed [O\,{\sc viii}]/[O\,{\sc vii}] intensity ratios within the Fermi bubbles  \citep{Sarkar_2023}.

Nuclear outflows from star formation in the Galactic center over long periods $\gtrsim30\,\mathrm{Myr}$ are also a possible explanation for the Fermi bubbles \citep{2011PhRvL.106j1102C, 10.1093/mnrasl/slu107, Crocker_2015, 10.1093/mnras/stz143, sands2025fermibubblesagngammaray}.

The nature of the eROSITA bubbles and their relationship to the Fermi bubbles remains ambiguous. \citet{Yang2022} proposes that both pairs of bubbles can be explained by a single jet-driven model. \citet{Mou2023} claims that they originate from distinct AGN wind-driven events. \citet{Zhang2024} proposes that the eROSITA bubbles are associated with the ring of star-forming clumps $\sim3\mathrm{-}5\,\mathrm{pc}$ from the GC. The northern eROSITA bubble hosts a bright arc feature known as the North Polar Spur/Loop I, discovered decades earlier in radio emission \citep{Tunmer1958ONAF}. Recent studies find morphological evidence that it is also of GC origin (e.g. \citealt{10.1093/mnras/stad1985, Liu_2024}).

Despite the deep characterization of the eROSITA and Fermi bubbles in the $\gamma$-ray, X-ray, and radio domains, very little work has focused on optical. \citet{Krishnarao_2020} reported the discovery of high-velocity H$\alpha$ emission associated with the Fermi bubbles using the Wisconsin H-Alpha Mapper (WHAM). \citet{10.1093/mnras/stad1985} further used WHAM to explore the NPS, finding a non-detection at the survey's depth limits. Here, I present the discovery of H$\alpha$, [O\,{\sc iii}], and [S\,{\sc ii}] filaments in the Northern Sky Narrowband Survey \citep{Ziegenbalg_2025}, coincident with high-resolution 144 MHz radio data from LoTSS DR3 \citep{Shimwell_2017, Shimwell_2026}. 

\section{Data}

\subsection{Northern Sky Narrowband Survey}

Optical emission line data in H$\alpha$, [O\,{\sc iii}], and [S\,{\sc ii}] are obtained from the recent Northern Sky Narrowband Survey (NSNS) DR0.2 \citep{Ziegenbalg_2025}. The narrowband observations were obtained using 5--10 cm aperture telescopes in bandpass widths of $35\,\mathrm{\AA}$ for H$\alpha$ $6563\,\mathrm{\AA}$ and $40\,\mathrm{\AA}$ for [O\,{\sc iii}] $5007\,\mathrm{\AA}$ as well as [S\,{\sc ii}] $6717,\,6730\,\mathrm{\AA}$. Visual continuum data were obtained in the SDSS $g$, $r$, and $i$ bands. For this work, a $50\degr\times100\degr$ cutout is obtained as the analysis field centered at $l=17.3\degr,\,b=44.6\degr$. The DR0.2 stacks are photometrically calibrated \citep{Roeser_2010} and converted to Rayleighs by calibrating to WHAM \citep{Haffner_2003}. A further calibration for [N\,{\sc ii}] variation is not applied in this work; while NSNS is zero-point calibrated to WHAM, the additional intensity ratio of NSNS/WHAM is consistent to within $\lesssim5\%$ at all $b$ values, not carrying the general increase in forbidden line ratios with $b$ and thus not requiring a correction.

\subsection{Radio Surveys}

Structure in the optical images is compared to the temperature-calibrated $408\,\mathrm{MHz}$ Haslam map \citep{1981A&A...100..209H}, which detects the wide scale North Polar Spur structure. Data from the LOw-Frequency-ARray (LOFAR) Two-metre Sky Survey in 144 MHz \citep{Shimwell_2017}, at a higher resolution of $\sim6$ arcsec, are used to analyze small-scale filamentary structure in the optical images.

\section{Methods}

\begin{figure*}
    \centering
    \includegraphics[width=1\linewidth]{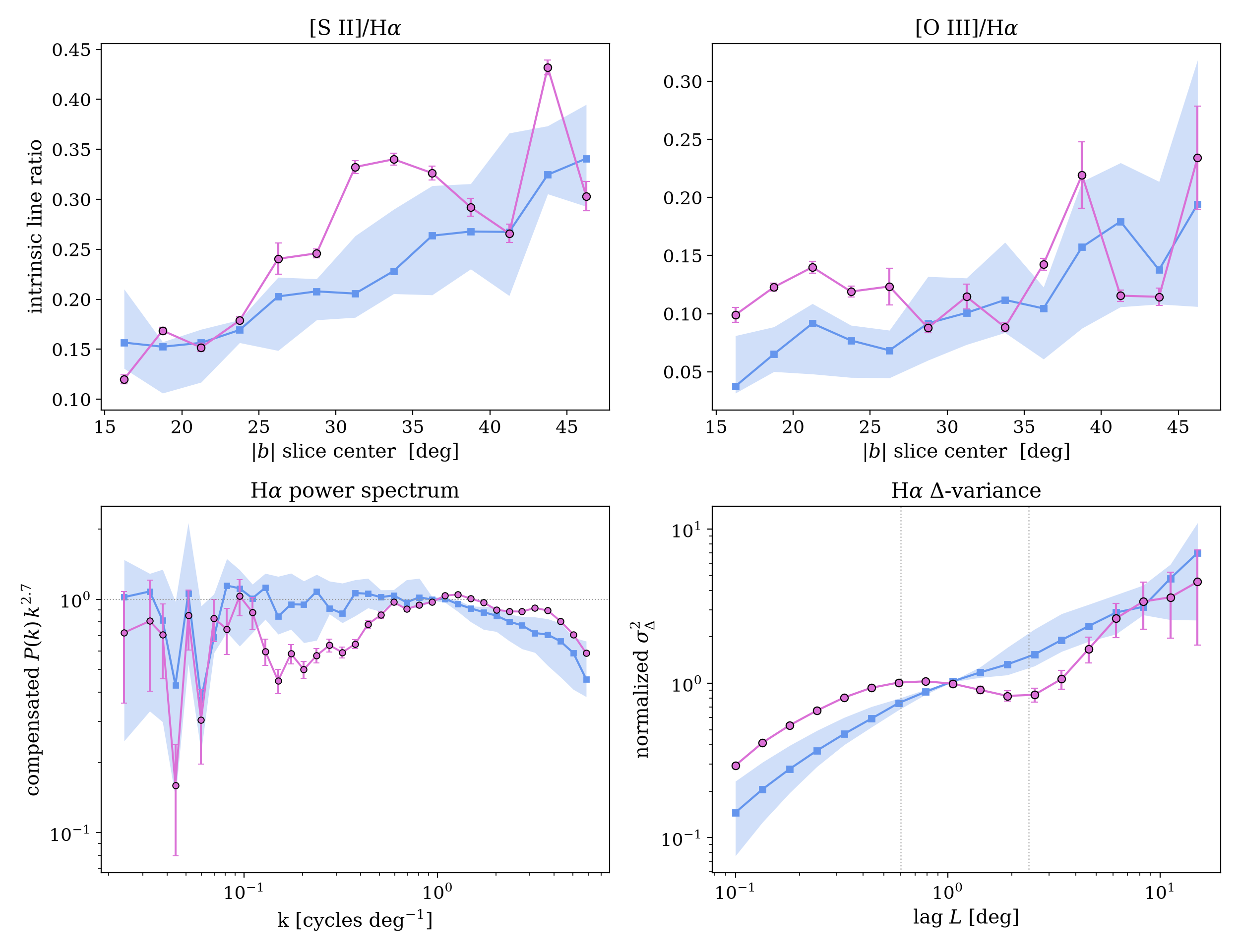}
    \caption{$\text{[S\,{\sc ii}]}/\mathrm{H}\alpha$ (top left) and $\text{[O\,{\sc iii}]}/\mathrm{H}\alpha$ (top right) line ratios of the NPS (purple) and comparison WIM regions (blue) by Galactic latitude. The H$\alpha$ power spectrum and $\Delta$-variance are displayed on the bottom left and right, respectively. Error bars are statistical uncertainty, and do not include an additional $\sim15\%$ systematic uncertainty.}
    \label{fig:filaments}
\end{figure*}

\begin{figure*}
    \centering
    \includegraphics[width=1\linewidth]{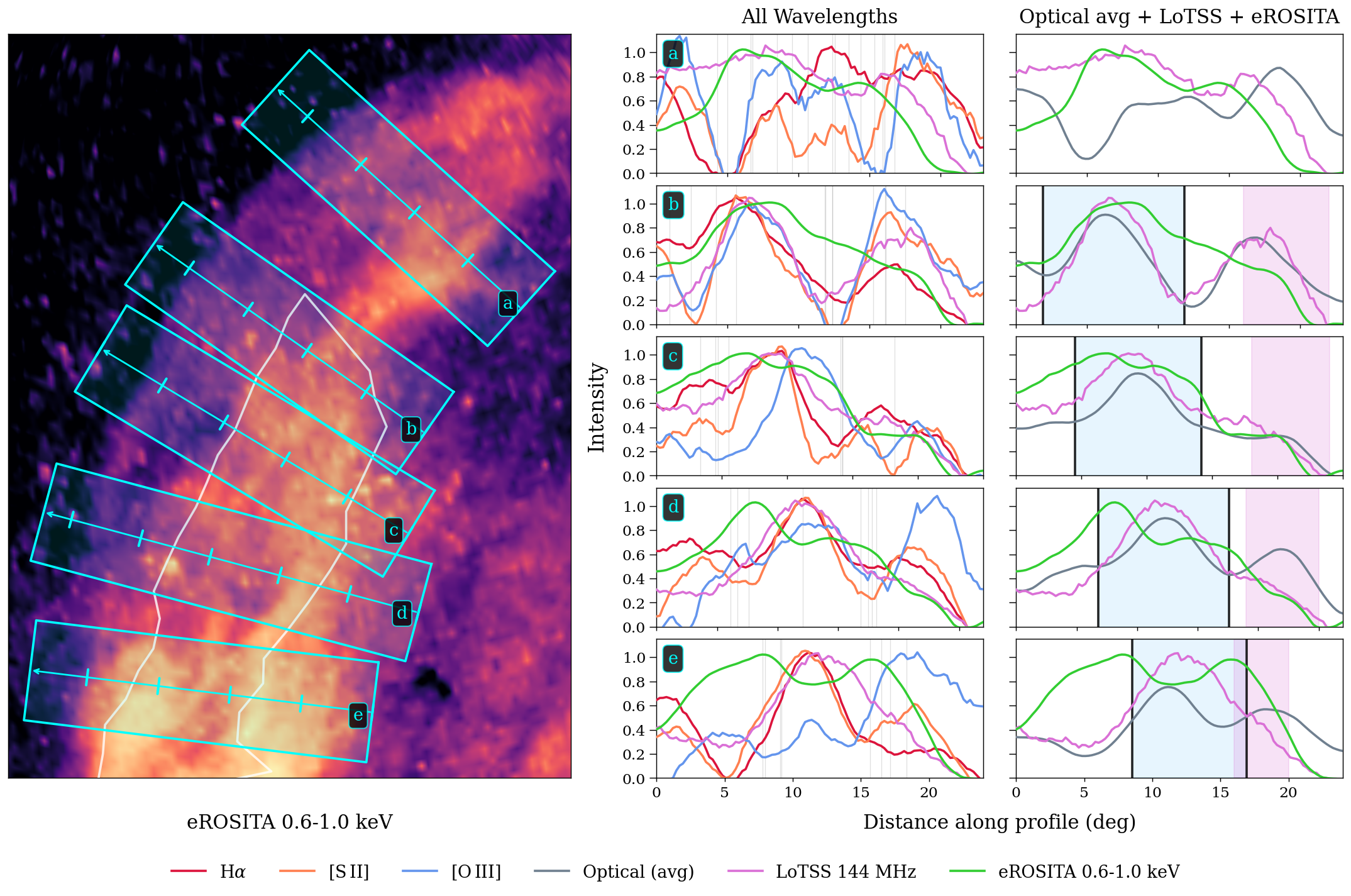}
    \caption{Multiwavelength profiles (a-e) of the North Polar Spur in H$\alpha$, [O\,{\sc iii}], and [S\,{\sc ii}] (Northern Sky Narrowband Survey), $144\,\mathrm{MHz}$ (LoTSS), and $0.6\mathrm{-}1.0\,\mathrm{keV}$ (eROSITA, left image). The magnetized Galactic structure \citep{Zhang2024} is shown in white. The first column shows the intensities of all wavelengths normalized to their peaks in the profiles. The right column displays the averaged optical data with LoTSS and eROSITA, and highlighted regions for the inner emission peak (blue) and eROSITA bubble edge (purple). All points from the magnetized structure are shown as faint gray lines in the left column, and averages of the clustered points are shown on the right as bold lines defining the inner region.}
    \label{fig:multiwavelength}
\end{figure*}

\subsection{Extinction Correction}

Given the eROSITA bubble/NPS's proximity to the Galactic plane, extinction is a critical consideration for assessing emission fluxes. The Bayestar19 dust maps by \citet{Green_2019} are used through the \texttt{dustmaps} package \citep{Green2018}, and the \citet{2019ApJ...886..108F} extinction law is applied per-pixel on the optical data. The Bayestar19 maps are distance-resolved and show no excess reddening at any distance along the NPS mask, indicating that the filaments are not associated with foreground dust such as the Aquila Rift, visible at low $b$ in the $E(B-V)$ panel of Figure \ref{fig:maps}. The nearby \ion{H}{1} structure at $\lesssim 140\,\mathrm{pc}$ described by \citet{10.1093/mnras/staa2702} is also visible in the $E(B-V)$ map at greater Galactic longitudes and does not coincide with the optical detection.

\subsection{Structure Masks}

To isolate and analyze the physical properties of the eROSITA filaments, a series of structure masks is created. A Gaussian smoothing filter with $\sigma=400\,\mathrm{pixels}$  is applied to the 408 MHz image in the analysis field. The residual Haslam image, obtained by subtracting the Gaussian smoothed image from the raw Haslam image, is used as a mask on the optical data to restrict the signal to the projection region of the NPS.

The NSNS images are subject to low frequency background grid artifacts in the [O\,{\sc iii}] and [S\,{\sc ii}] channels \citep{Ziegenbalg_2025}. To minimize detections of these artifacts, a consensus mask is defined. In each Gaussian-subtracted channel, emission is masked, and pixels are used where they exceed $2.5\sigma$ of the local background noise. The integrated $S/N$ of the emission in each filter mask is $\gtrsim65$. The consensus mask contains regions where these masks in H$\alpha$ intersect the masks in either [O\,{\sc iii}] or [S\,{\sc ii}]:
\begin{equation}
M_\mathrm{consensus,\,2.5\sigma}=M_\mathrm{H\alpha,\,2.5\sigma}\cap (M_\mathrm{[O\,{\sc iii}],\,2.5\sigma}\cup M_\mathrm{[S\,{\sc ii}],\,2.5\sigma})
\end{equation}
A similar mask is obtained to extract filaments in the LoTSS 144 MHz data. To select extragalactic sources, a query is performed on the VO @ ASTRON TAP service by applying a selection threshold of total flux $>30\,\mathrm{mJy}$. Each selected source is expanded by a 4-pixel radius and subtracted from the image. A Gaussian smoothing filter of $\sigma=80\,\mathrm{pixels}$ is applied to the extragalactic source-removed LoTSS data. The resulting residual image from subtracting the smoothed image is used as the corresponding LoTSS filament mask.

\subsection{Filament Properties}

As the H$\alpha$ sky contains many diffuse structures extending from the Galactic plane, such as the warm ionized medium (WIM; e.g. \citealt{RevModPhys.81.969}), it is important to distinguish WIM filaments from emission associated with the NPS. Eight comparison regions are selected using the same FOV and consensus mask methodology as the NPS panel at differing Galactic longitudes, and are used as control measurements of the WIM and general Galactic outflows. For consistency, the Haslam mask is not used in this analysis, and fixed $b$ slices are instead measured.

The WIM is observed to increase in $\text{[N\,{\sc ii}]}/\mathrm{H}\alpha$, $\text{[S\,{\sc ii}]}/\mathrm{H}\alpha$ and $\text{[O\,{\sc iii}]}/\mathrm{H}\alpha$ with decreasing H$\alpha$ intensities \citep{1999ApJ...523..223H, RevModPhys.81.969}, and thus broadly increases with Galactic latitude. The relationship is observed in the NSNS $\text{[S\,{\sc ii}]}/\mathrm{H}\alpha$ and $\text{[O\,{\sc iii}]}/\mathrm{H}\alpha$ for the WIM comparison regions, shown in blue in the top two panels of Figure \ref{fig:filaments}. The NPS, observed in purple, diverges from the relation with greater $\text{[S\,{\sc ii}]}/\mathrm{H}\alpha$ at $\sim25\mathrm{-}40\degr$, indicating a potential shock or photoionization component distinct from the WIM. The $\text{[O\,{\sc iii}]}/\mathrm{H}\alpha$ of the NPS may be elevated at low-$b$, but is otherwise not distinct.

The electron density spectrum in the interstellar medium follows a remarkably smooth, well-studied Kolmogorov power law, known as the ``big power law in the sky'' (e.g. \citealt{1995ApJ...443..209A}). \citet{Chepurnov_2010} find consistency between the Kolmogorov spectrum and the WHAM H$\alpha$ survey data. The H$\alpha$ power spectrum of the warm ionized medium should thus follow a similarly smooth spectral slope. The $\Delta$-variance method evaluates the power spectrum by measuring structure on a given scale $l$ in a map $f(r)$ by filtering the map with a Mexican-hat filter ($\bigodot\nolimits_{l}$) and computing its variance \citep{1966IEEEP..54..221A, 1998A&A...336..697S, Bensch_2001, 2008A&A...485..917O, Ossenkopf_2008b}:
\begin{equation}
   \sigma^2_\Delta(l) = \left\langle \left( f(\mathbf{r}) * \bigodot\nolimits_{l}(\mathbf{r}) \right)^2 \right\rangle_{\mathbf{r}}
\end{equation}

\citet{Elia_2014} measures the $\Delta$-variance of diffuse infrared Galactic emission, finding that the diffuse structure generally follows a fractional Brownian motion curve behavior, while compact sources are responsible for bumps. The H$\alpha $ $\Delta$-variance is calculated for the NPS and comparison WIM fields. The WIM follows the expected featureless curve, whereas the NPS demonstrates a clear break, peaking at $\mathrm{lag}\,L\approx0.6\degr$, shown in the bottom row of Figure \ref{fig:filaments}. The scale is consistent with nondetection in WHAM, which is limited by its beam angular resolution of $1\degr$.

\section{Multiwavelength Analysis}

\subsection{Spatial Coincidence}

Analyzing the relationship between the NPS optical emission and radio and X-ray data is critical for testing their association. Five profiles are placed along the North Polar Spur at varying Galactic latitudes and are angled perpendicularly to the edge, labeled a-e in Figure \ref{fig:multiwavelength}.

\citet{Zhang2024} identified several magnetized structures emerging from the Galactic center through analysis of the polarized synchrotron emission, potentially associated with the eROSITA bubbles. Profiles b-e align with two of these structures. Within these profiles, the NSNS optical emission shows clear spatial coincidence with the LoTSS $144\,\mathrm{MHz}$ emission as an intensity peak on the interior of the NPS, displayed as the blue highlighted region in the right column of Figure \ref{fig:multiwavelength}. This inner region also demonstrates a relationship with the magnetized structure, which concentrates on either edge of the intensity bump. The alignment may be a result of magnetic draping surrounding the structure. Profile a, which lies outside of the magnetized region, does not detect the inner region in either NSNS or LoTSS. Fainter emission is observed at the edge of the eROSITA bubble, highlighted in purple, in both optical and $144\,\mathrm{MHz}$.

The H$\alpha$ and [S\,{\sc ii}] emission generally follow similar normalized intensity distributions within the magnetized region. The [O\,{\sc iii}] exhibits a far more variable distribution, and has relatively prominent structure at the outer bubble edge, though the data also contain more noise.

Only a low-resolution image of the eROSITA $0.6\mathrm{-}1.0\,\mathrm{keV}$ data is available from \citet{Predehl2020}, which does not show a clear correlation with optical or $144\,\mathrm{MHz}$ in any of the profiles.

\subsection{Intensity and Temperature}

The strong spatial coincidence of structure in H$\alpha$, [S\,II], and [O\,III] with 144 MHz supports a distinct shock ionized or photoionized source rather than WIM. To determine whether the associated NPS radio emission could be from free-free radiation, the expected brightness temperature is derived from the H$\alpha$.

The intensity of H$\alpha$ is measured to be $3.07\pm1.04\,\mathrm{R}$ within the consensus mask and $1.98\pm0.36\,\mathrm{R}$ when confined to the bounds of the magnetized structure. Using the OFF-NPS H$\alpha$ background measurement of $1.22\,\mathrm{R}$ from \citet{10.1093/mnras/stad1985}, the excess intensity of the consensus mask is $\Delta I_{\mathrm{H}\alpha}=1.85\pm1.04\,\mathrm{R}$.

The expected absorption from free-free is described as
\begin{equation}
a^{ff}_v=0.018T^{-3/2}Z^2n_en_iv^{-2}\bar{g}_{ff}
\end{equation}
by \citet{RevModPhys.81.969} in $\mathrm{cm}^{-1}$. Using the velocity averaged Gaunt factor
\begin{equation}
    \bar{g}_{ff}(v,\,T_e)\approx\ln
\left\{
\exp\left[
5.960-\frac{\sqrt3}{\pi}\ln(Z_iv_9T^{-3/2}_4)
\right]+e
\right\}
\end{equation}
by \citet{2011piim.book.....D}, solving to obtain $\bar{g}_{ff}=6.28$, and case B recombination
\begin{equation}
I(\mathrm{H}\alpha)=9.41\times10^{-8}T^{-1.017}_410^{-0.029/T_4}(\mathrm{EM})
\end{equation}
from \citet{Reynolds2011}, the emission measure is $\mathrm{EM}\approx 2.25I_\mathrm{H\alpha}=4.2\pm2.3\,\mathrm{cm}^{-6}\,\mathrm{pc}$, assuming the typical WIM temperature $T_{e,\mathrm{WIM}}\approx8000\,\mathrm{K}$. The optical depth $\tau_v$ is calculated using
\begin{equation}
\tau_v=1.772\times10^{-26}T^{-3/2}_4v^{-2}_9\bar{g}_{ff}\left(\frac{n_i}{n_p}\right)\mathrm{EM}
\end{equation}
\citep{2011piim.book.....D}, to find $\tau_v=(1.2\pm0.7)\times10^{-5}$. This is applied to the brightness temperature:
\begin{equation}
    T_b=T_e(1-e^{-\tau})
\end{equation}
by \citet{condon2016essential}, considering a $(1+0.08)$ contribution factor from $\mathrm{He}$ \citep{10.1046/j.1365-8711.2003.06439.x}, resulting in the free-free brightness temperature $T_b\approx0.1\,\mathrm{K}$. This is far below the observed temperature of Haslam $T_{b,\,408\,\mathrm{MHz}}=7.3\pm4.2\,\mathrm{K}$, calibrated by the OFF-NPS background of $35\,\mathrm{K}$ from \citet{10.1093/mnras/stad1985}. Thus, the observed radio brightness of the NPS is not explained by free-free radiation.

\citet{10.1093/mnras/stad1985} defines a parameter $\mathcal{Q}=I_\mathrm{H\alpha}/I_{1.4\,\mathrm{GHz}}$. The intensity $I_{1.4\,\mathrm{GHz}}$ is found from the Rayleigh--Jeans approximation $I_{v}={2kT_bv^2}/{c^2}$ to obtain $I_{1.4}=(2.3\pm1.3)\times10^{-10}\,\mathrm{erg}\,\mathrm{cm}^{-2}\,\mathrm{s}^{-1}\,\mathrm{sr}^{-1}$. Measured across the full Haslam mask, $\mathcal{Q}_\mathrm{NPS}\approx320\mathrm{-440}$, which is above the limit $\mathcal{Q}_\mathrm{NPS,\,WHAM}\lesssim50$ from \citet{10.1093/mnras/stad1985}, as the filaments are unresolved in WHAM. The observed value remains far below that of typical SNRs $\mathcal{Q}\approx10^4$, favoring a Galactic center nature of the NPS over a local shock. The observed H$\alpha$ from a local ionized shell would require an intensity $\sim25$ times greater.

\citet{Krishnarao_2020} detected high-velocity H$\alpha$ emission towards the Fermi bubbles using WHAM, measuring $I_{\mathrm{H}\alpha}=0.84^{+0.10}_{-0.09}\,\mathrm{R}$ and $EM=2.00^{+0.64}_{-0.63}\,\mathrm{cm^{-6}}\,\mathrm{pc}$ for the kinematically isolated component, comparable to the values in this work, with the factor $\sim2$ difference potentially reflecting the greater line of sight depth of the eROSITA bubble and distinct measurement techniques.

\begin{deluxetable}{lcc}
\setlength{\tabcolsep}{18pt}
\tabletypesize{\scriptsize}
\tablecaption{North Polar Spur Emission-Line Properties and Energetics\label{tab:energetics}}
\tablehead{
  \colhead{Parameter} &
  \colhead{Units} &
  \colhead{Value}
}
\startdata
$I_{\rm H\alpha}$        & R                    & $3.07\pm1.04$ \\
$I_\mathrm{[S\,II]}$     & R                    & $0.61\pm0.17$ \\
$I_\mathrm{[O\,III]}$    & R                    & $0.42\pm0.17$ \\
$\mathcal{Q}$            & ---                  & $\sim320\mathrm{-}440$ \\
$A_V$                    & mag                  & $0.33^{+0.27}_{-0.14}$ \\
{[\ion{S}{2}]}/H$\alpha$ & ---                  & $\sim0.12\mathrm{-}0.43$ \\
{[\ion{O}{3}]}/H$\alpha$ & ---                  & $\sim0.09\mathrm{-}0.23$ \\
$P_\mathrm{req}$ & $\mathrm{erg\,s}^{-1}$ & $5\mathrm{-}12\times10^{41}$
\enddata
\tablecomments{Intensity and line ratio values are not corrected for diffuse background. The background subtracted value for H$\alpha$ is $1.85\pm1.04$.}
\end{deluxetable}

\begin{figure*}
    \centering
    \includegraphics[width=1\linewidth]{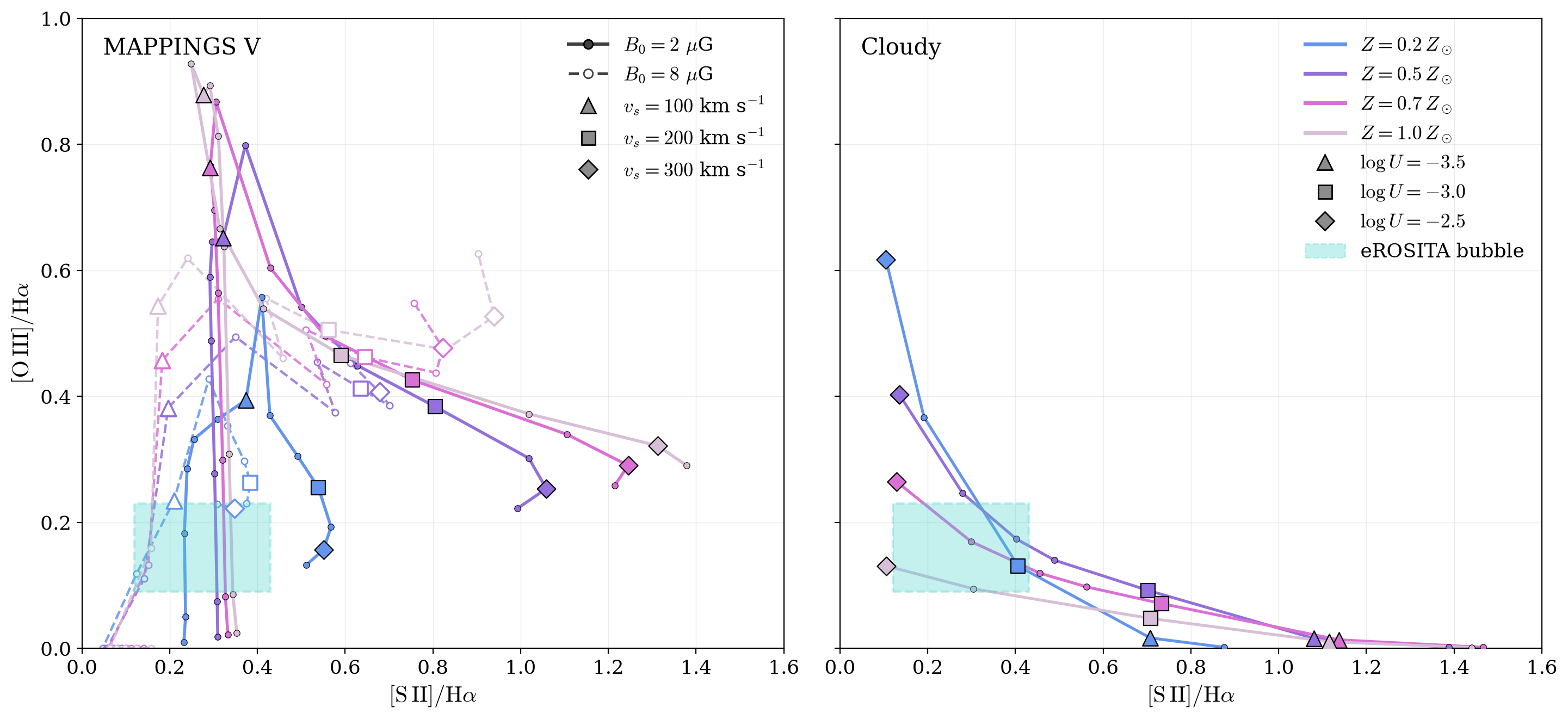}
    \caption{MAPPINGS V shock models (left) and Cloudy photoionization models (right) of the NPS. The lines are colorized by input metallicity $Z$ values. MAPPINGS V is run for two $B_0$ values $2\,\mu G$ and $8\,\mu G$, which is not well constrained for the NPS. The model curves vary significantly with their input $Z$ and $B$ parameters. Symbols represent the $v_s$ values for model points in MAPPINGS V and the ionization parameter $\log U$ in Cloudy. The full range of NSNS measurement ratios of the NPS in the eROSITA bubbles is shown in cyan.}
    \label{fig:models}
\end{figure*}

\subsection{Energetics}

The electron density is estimated from the H$\alpha$ $EM$
\begin{equation}
    n_e\approx\sqrt{EM/fL}
\end{equation}
where $f\sim0.4$ is the filling fraction and $L\sim1\,\mathrm{kpc}$ is the line of sight depth. The ionized mass is determined using the relation \citep{Ciampa_2021} adopting a GC origin
\begin{equation}
    M_\mathrm{ion}=\mu n_e\Omega D^2L\cos(i)
\end{equation}
where $\Omega\sim0.12\,\mathrm{sr}$ is the solid angle of the consensus mask, $D\sim6.5\,\mathrm{kpc}$ is the heliocentric distance to the emission, $i\sim0$ is the sheet inclination, and $\mu=1.4\,\mathrm{m_H}$ is the mean gas mass per electron, to obtain $M_\mathrm{ion}\approx1.6\times10^7\,M_\odot$. The ionizing photon flux is defined as
\begin{equation}
    \phi_\mathrm {ion}=\alpha_Bn_eN_i=\alpha_BEM
\end{equation}
\citep{Bland-Hawthorn_1999}, which is multiplied by $\Omega D^2$ to obtain $Q_\mathrm{ion}\approx1\times10^{51}\,\mathrm{s^{-1}}$ necessary to sustain $M_\mathrm {ion}$. The required luminosity is thus $L_\mathrm{ion}=\langle E\rangle Q_\mathrm{ion}$. Assuming $\langle E\rangle\approx20\,\mathrm {eV}$ (e.g. \citealt{2011piim.book.....D}), $L_\mathrm{ion}\approx1.3\times10^{40}\,\mathrm{erg\,s}^{-1}$.

The power required from a given eROSITA bubble model for the observed luminosity is described as
\begin{equation}
P_\mathrm{req}=\frac{L_\mathrm{ion}}{\epsilon}\times \frac{4\pi r^2_\mathrm{GC}}{\Omega D^2}
\end{equation}
where $\epsilon$ is the efficiency of source power $P_\mathrm{req}$ in ionizing the NPS. Adopting $\epsilon=0.5$ as a conservative, high-efficiency boundary and $r_\mathrm{GC}\approx4\mathrm{-}6.5\,\mathrm{kpc}$, $P_\mathrm{req}=5\mathrm{-}12\times10^{41}\,\mathrm{erg\,s}^{-1}$.

The observed $P_\mathrm{req}$ disfavors the proposed power of star formation driven wind models $\sim5\times10^{40}\,\mathrm{erg\,s}^{-1}$ (e.g. \citealt{10.1093/mnras/stv1806}), which is insufficient even with perfect efficiency ($\epsilon=1.0$). AGN jet models predict powers $\sim10^{44}\,\mathrm{erg\,s}^{-1}$ (e.g. \citealt{Yang2022}), which comfortably exceeds the required ionizing power of the NPS by $\sim2$ orders of magnitude. Hot accretion AGN flow models have values $\sim2\times10^{41}\mathrm{erg\,s}^{-1}$ \citep{Mou_2014}, and are not excluded. The star formation ring model by \citet{Zhang2024} proposes the outflow being driven by star-forming clumps $\sim3\mathrm{-}5\,\mathrm{pc}$, shifting $P_\mathrm{req}=1.2\mathrm{-}5\times10^{41}\,\mathrm{erg\,s}^{-1}$ for the NPS. The injection rate $\sim1\mathrm{-}9\times10^{40}\,\mathrm{erg\,s}^{-1}$ is lower than $P_\mathrm{req}$, but not clearly excluded. The spatial alignment of the magnetic structure around the inner NPS emission makes the SF ring model particularly compelling. The brightest clumps in the Hi-GAL star formation map \citep{Elia_2022} are also coincident with the base of the NPS.

\subsection{Models}

To simulate potential shock ionization and photoionization sources of the optical cloud, models are created using MAPPINGS V \citep{2018ascl.soft07005S} and Cloudy \citep{1998PASP..110..761F, Gunasekera_2023, 2025RMxAA..61c.120G}, respectively, provided in Figure \ref{fig:models}. MAPPINGS V shocks are computed for the parameters $n_0=1.8\,\mathrm{cm}^{-3}$, $v_s=60\mathrm{-}350\,\mathrm{km\,s}^{-1}$, $B_0=2\,\mu G$ and $B_0=8\,\mu G$, and across a metallicity range $Z=0.2\mathrm{-}1.0\,\mathrm{Z}_\odot$. The Cloudy photoionization models are created under the same metallicity range for a static sphere illuminated by a $4\times10^4\,\mathrm{K}$ blackbody, with a density $n_H=0.1\,\mathrm{cm}^{-3}$, and across a range of ionization parameters $\log U=-4.0$ to $-2.5$.

Due to the lack of precise measurements of $Z$ and $B$ for this region in the literature, the uncertainty of $\text{[S\,{\sc ii}]}/\mathrm{H}\alpha$ and $\text{[O\,{\sc iii}]}/\mathrm{H}\alpha$ over the NPS, and a lack of subtraction of ambient WIM, a specific ionization source cannot yet be determined. A deep follow-up program is required to measure the line ratios to spatially map the ionization properties of the NPS.

\section{Conclusion}

In this paper, I present the discovery of coincident optical and $144\,\mathrm{MHz}$ emission towards the North Polar Spur (NPS) and the northern eROSITA bubble. The signal is identified in H$\alpha$, [O\,{\sc iii}], and [S\,{\sc ii}] from the Northern Sky Narrowband Survey. To distinguish it from the typical warm ionized medium (WIM) at high Galactic latitudes, $\text{[S\,{\sc ii}]}/\mathrm{H}\alpha$ and $\text{[O\,{\sc iii}]}/\mathrm{H}\alpha$, power spectra, and $\Delta$-variance of the NPS are measured in comparison to background regions. The forbidden line ratios hint at potential enhancement relative to the smooth WIM relation with $b$. The H$\alpha$ power spectra and $\Delta$-variance provide a clear structural preference in the NPS for filaments at lag $L\approx0.6\degr$, breaking from the WIM relation. Profiles are measured across five regions in the NPS, indicating strong coincidence between the optical and LoTSS emission in a prominent inner region and a faint peak at the eROSITA bubble edge. Literature maps of the polarized synchrotron emission, representative of Galactic magnetized structure, cluster at the edges of the inner NPS region. The excess intensity of the H$\alpha$ is computed as $I_{\mathrm{H\alpha}}=1.85\pm1.04\,\mathrm{R}$, following background subtraction. The faintness of H$\alpha$ relative to Haslam $408\,\mathrm{MHz}$ in the NPS is incompatible with a local SNR-like shock. Assuming the NPS is of a GC origin, analysis of the required ionization power of the NPS H$\alpha$ yields $P_\mathrm{req}=5\mathrm{-}12\times10^{41}\,\mathrm{erg\,s}^{-1}$, disfavoring star-formation wind models of the eROSITA bubble regardless of efficiency. AGN jets and accretion flow models, as well as the SF ring, are in broad agreement with H$\alpha$ intensity.

\begin{acknowledgments}

I am grateful to Patrick Ogle, Mary Putman, and David Schiminovich for insightful conversations and helpful comments. I thank Stefan Ziegenbalg for the publicly accessible DR0.2 of the Northern Sky Narrowband Survey, which is essential to this work.

LOFAR is the Low Frequency Array designed and constructed by ASTRON. It has observing, data processing, and data storage facilities in several countries, which are owned by various parties (each with their own funding sources), and which are collectively operated by the LOFAR ERIC under a joint scientific policy. The LOFAR resources have benefited from the following recent major funding sources: CNRS-INSU, Observatoire de Paris and Université d'Orléans, France; BMFTR, MKW-NRW, MPG, Germany; Science Foundation Ireland (SFI), Department of Business, Enterprise and Innovation (DBEI), Ireland; NWO, The Netherlands; The Science and Technology Facilities Council, UK; Ministry of Science and Higher Education, Poland; The Istituto Nazionale di Astrofisica (INAF), Italy.  

This work is based on data from eROSITA, the soft X-ray instrument aboard SRG, a joint Russian-German science mission supported by the Russian Space Agency (Roskosmos), in the interests of the Russian Academy of Sciences represented by its Space Research Institute (IKI), and the Deutsches Zentrum für Luft- und Raumfahrt (DLR). The SRG spacecraft was built by Lavochkin Association (NPOL) and its subcontractors, and is operated by NPOL with support from the Max Planck Institute for Extraterrestrial Physics (MPE). The development and construction of the eROSITA X-ray instrument was led by MPE, with contributions from the Dr. Karl Remeis Observatory Bamberg \& ECAP (FAU Erlangen-Nuernberg), the University of Hamburg Observatory, the Leibniz Institute for Astrophysics Potsdam (AIP), and the Institute for Astronomy and Astrophysics of the University of Tübingen, with the support of DLR and the Max Planck Society. The Argelander Institute for Astronomy of the University of Bonn and the Ludwig Maximilians Universität Munich also participated in the science preparation for eROSITA.

The Wisconsin H$\alpha$ Mapper and its H$\alpha$ Sky Survey have been funded primarily by the National Science Foundation. The facility was designed and built with the help of the University of Wisconsin Graduate School, Physical Sciences Lab, and Space Astronomy Lab. NOAO staff at Kitt Peak and Cerro Tololo provided on-site support for its remote operation.

\end{acknowledgments}

\software{Astropy \citep{astropy:2013, astropy:2018, astropy:2022}, dustmaps \citep{Green2018}, MAPPINGS V \citep{2018ascl.soft07005S}, Cloudy \citep{1998PASP..110..761F,Gunasekera_2023, 2025RMxAA..61c.120G}, NumPy \citep{harris2020array}, SciPy \citep{2020SciPy-NMeth}, Matplotlib \citep{Hunter:2007}}

\bibliographystyle{aasjournalv7}
\bibliography{references}



\end{document}